\documentclass[twoside]{dis07}
\usepackage[latin1]{inputenc}
\usepackage[dvips]{graphicx,epsfig,color}
\usepackage{wrapfig,rotating}
\usepackage{amssymb,amsmath,array}
\usepackage{cite}

\def\kt{\ensuremath{k_\perp}}
\def\pt{\ensuremath{p_t}}

\newcommand{\as}{{\alpha}_\mathrm{s}}

\newcommand{\Pmax}{\bar{q}}

\begin{document}
\title{Towards precision determination of uPDFs}

\author{Magnus Hansson$^1$ and Hannes Jung$^2$
%
%
\vspace{.3cm}\\
%
1- Lund University
%
\vspace{.1cm}\\
2- DESY, FRG
}

\maketitle

\begin{abstract}
The unintegrated Parton Density
Function of the gluon is obtained  from a fit to 
 dijet production in DIS as measured at HERA. Reasonable descriptions
of the measurements are obtained, and a first attempt to constrain the intrinsic transverse momentum
distribution at small $\kt$ is presented~\cite{url}.
\end{abstract}
\section{Introduction}
\begin{wrapfigure}{r}{0.5\columnwidth}
\centerline{\includegraphics[width=0.45\columnwidth]{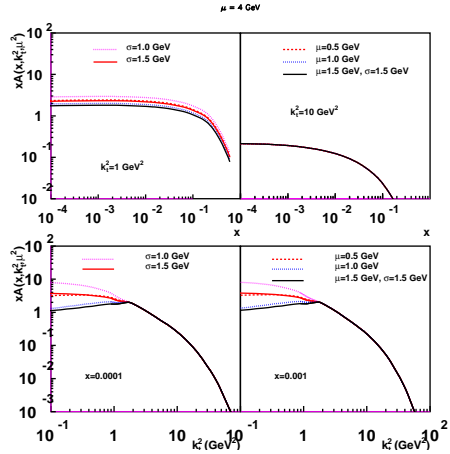}}
\caption{The unintegrated gluon distribution at a scale $\Pmax=4$~GeV for different 
values of $\mu$ and $\sigma$ of the intrinsic $\kt$ distribution
as a function of $x$ for fixed $\kt$(top) and as a function of $\kt$ (bottom)
for fixed $x$ }
\label{intrinsic-kt}
\end{wrapfigure}
Unintegrated 
parton density functions (uPDFs) 
are best suited to study details of the
hadronic final state in high energy $ep$ and also in $pp$ collisions (for a
review see
\cite{jung-hq-2001,jung-mpla2003,smallx_2001,smallx_2002,
smallx_2004,jung-dis04,jung-collins-2005}). 
In general, the production 
cross section for jets, heavy quarks or gauge bosons
can be
written as a convolution of the uPDF ${\cal A}(x,\kt^2,\Pmax)$ with the 
partonic off-shell cross section ${\hat\sigma}(x_i, \kt^2)$, with 
$x_i, \kt$ being the longitudinal momentum fraction and the transverse momentum
of the interacting parton $i$ and $\Pmax $ being the factorization scale. For
example the cross section for  $ep \to \mbox{jets} + X $ can be written as:
\begin{eqnarray*}
\frac{d \sigma^{jets}}{d E_T d \eta} &=&
\sum_i \int\int\int dx_i\; dQ^2 d\dots \\
& & \cdot \left[ d\kt^2 x_i{\cal A}(x_i,\kt^2,\Pmax)\right] {\hat\sigma} (x_i, \kt^2)
 \end{eqnarray*}
At high energies, the gluon density
 is dominating for many processes, therefore 
here only the gluon uPDF is considered. It has already been
shown in \cite{heralhc2006}, that the predictions 
of the total cross section as well as  
differential distributions for heavy quark production at HERA
and the LHC agree well in general with those coming from fixed NLO 
calculations. However, the details
depend crucially on a precise knowledge of the uPDF. Therefore precision fits to
inclusive and exclusive measurements have to be performed to determine precisely
the free parameters of the uPDF: the starting distribution
function at a low scale $Q_0 \sim 1$ GeV as well as parameters connected
with $\as$ and details of the splitting functions for the perturbative evolution. 

An overview and discussion of uPDFs is given in \cite{smallx_2001,smallx_2002,
smallx_2004}. 
In a previous paper~\cite{jung-ichep06} the uPDF was determined from a pQCD fit using the CCFM
evolution equation \cite{CCFMa,CCFMb,CCFMc,CCFMd} to the structure function 
$F_2$ and $F_2^c$ with acceptable $\chi^2/ndf $. However, the small $x$ behavior of the uPDF obtained
from $F_2^c$ was very different compared to the one obtained from $F_2$. 

Here also measurements of high $\pt$-dijet production in DIS 
at HERA~\cite{Romans-jets,h1-delta-phi,zeus-delta-phi} are investigated.
\section{The method}
The
unintegrated gluon density is determined by a convolution of the
non-perturbative starting distribution ${\cal A}_0 (x)$ and
the CCFM evolution denoted by 
${\cal \tilde A}\left(x,\kt,\Pmax\right)$: 
\begin{eqnarray*}
x {\cal A}(x,\kt,\Pmax) &= &\int dx' 
{{\cal A}_0 (x',\kt) }\cdot \frac{x}{x'}{ {\cal \tilde A}\left(\frac{x}{x'},\kt,\Pmax\right) }\nonumber 
\end{eqnarray*}
In the perturbative evolution  the  gluon splitting function $P_{gg}$ 
including non-singular terms 
(as described in detail in \cite{jung-dis02,jung-dis03}) is applied.
\begin{wrapfigure}{r}{0.5\columnwidth}
\centerline{\includegraphics[width=0.45\columnwidth]{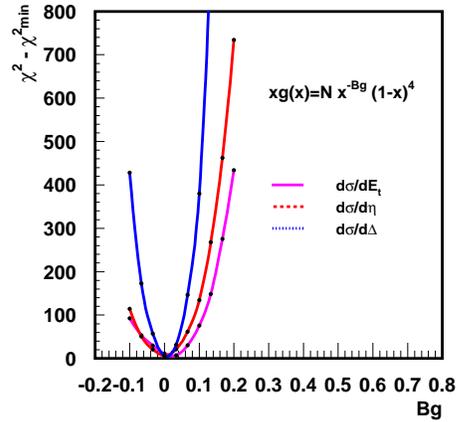}}
\caption{A scan in the parameter space of $B_g$ for
$\frac{d^3\sigma }{dQ^2 dx d E_t}$, $\frac{d^3\sigma}{dQ^2 dx d \Delta }$ 
and $\frac{d^3\sigma}{ dQ^2 dxd \Delta \eta}$ as
measured in 
\protect\cite{Romans-jets}. }
\label{jet-scan}
\end{wrapfigure}

The distribution ${\cal A}_0$ is parameterized at the starting scale $Q_0$
by:
\begin{eqnarray}
x{\cal A}_0(x,\kt) &=& N x^{-B_g} \cdot (1 -x)^{C_g}\left( 1 -D_g x\right) 
\nonumber \\
& & \cdot \exp\left[ 
-\frac{(\mu-\kt)^2}{\sigma^2}\right]
\label{a0}
\end{eqnarray}
The parameters $N_g, B_g, C_g, D_g$ as well as $\mu, \sigma$ of ${\cal A}_0$ are free parameters
which have to be constrained by measurements.
 It turns out, that $C_g, D_g$ are not sensitive to the data considered here, and are therefore
fixed to $C_g=4$ and $D_g=0$. The other parameters are determined by a fit~\cite{minuit} to 
measurements
such to minimize the $\chi^2$ defined by:
\begin{equation}
\chi^2 = 
\sum_i \left( \frac{\left(T - D  \right)^2}
{\sigma_i^{2\;stat} + \sigma_i^{2\;sys}}\right)  \nonumber 
\end{equation}
with $T $ being the theory value and $D$ the measurement with 
the corresponding statistical and systematic uncertainty.

\section{The intrinsic \kt distribution}
The Gaussian  form with $\mu=0$ and a width of $\sigma \sim 1.0$~GeV of the intrinsic 
\kt distribution in eq.(\ref{a0}) is an assumption to
parameterize our ignorance about the small \kt behavior. In the saturation model of
GBW~\cite{gbw} the uPDF vanishes for small \kt. Such a behavior can be mimicked by 
a Gaussian distribution with $\mu \sim Q_0$. The effect of choosing different $\mu$ is 
illustrated in  Fig.~\ref{intrinsic-kt}.

\section{Dijets in DIS}
The sensitivity of the shape in $x$ and the intrinsic \kt was
studied for dijets in DIS \cite{Romans-jets} in the kinematic
range of $5<Q^2<100$~GeV$^2$, $10^{-4}<x<10^{-2}$, $0.1 < y <0.7$
and two jets with at least $E_t > 5$ GeV in the range $-1<\eta<2.5$. The
differential cross
sections $\frac{d\sigma}{dE_t}$, $\frac{d\sigma}{d \Delta\eta}$, with $\Delta
\eta$ being the rapidity difference between the highest $E_t$ jets  
are mainly sensitive to the $x$ dependence of the 
uPDF. The same is observed for the cross section $\frac{d\sigma}{d \Delta}$  
with $E_t>E_{t\;min} +\Delta$
and $E_{t\;min} =5$~GeV.
 A scan over the 
parameter space of $Bg$ is shown in
Fig~\ref{jet-scan}. 
With this choice of parameters the cross sections are well described, giving a 
reasonable $\chi^2/ndf$.  
In Tab.~\ref{chi2} the $\chi^2/ndf$ are given for different values of $Bg$ and
the mean $\mu$ of the intrinsic $\kt$ distribution.
\begin{wraptable}{l}{0.7\columnwidth}
\centerline{\begin{tabular}{|l|c|c|c|c|}
\hline
 & & \multicolumn{3}{c|}{$\chi^2/ndf$} \\\hline
Bg &$\mu$ [GeV]  & $\frac{d\sigma}{dE_t}$ & $\frac{d\sigma}{d \Delta\eta}$ & 
$\frac{d\sigma}{d \Delta}$ \\\hline
 0.025  &  1.5    &  68/37=1.8  & 102/35=2.3  &  267/89=3.0 \\ \hline
 0.25   &  1.5    &  95/37=2.5  &  113/35=2.5  &  306/89=3.4 \\ \hline
 0.025  &  0   &  63/37=1.7  &  93/35=2.1  &  284/89=3.2 \\ \hline
 0.25   &  0    &  99/37=2.7  &  123/35=2.7  &  345/89=3.9 \\ \hline
\end{tabular}}
\caption{Quality of the description of the different differential cross sections
using $Bg=0.025$ and $Bg=0.25$ together with $\sigma=1.5$~GeV. }
\label{chi2}
\end{wraptable}

From Tab.~\ref{chi2} it is seen, that a value of $Bg=0.025$ is preferred, and
that the
sensitivity of these measurements 
to the intrinsic $\kt$ distribution is very small. 

However, the cross section as a function of $\Delta \phi$, where $\Delta\phi$ is the difference
in azimuthal angle between the two leading jets in the hadronic center-of-mass
frame, is directly sensitive to the
transverse momentum of the incoming parton, and thus a crucial test of the uPDF.

\begin{wrapfigure}{r}{0.5\columnwidth}
\centerline{\includegraphics[width=0.5\columnwidth]{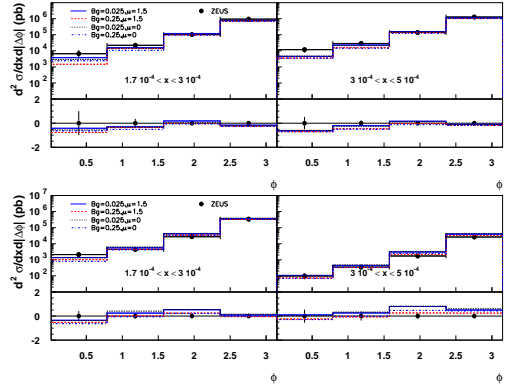}}
\caption{The cross section $\frac{d\sigma}{d\Delta \phi}$ as measured by 
\protect\cite{zeus-delta-phi} compared to predictions using CASCADE and the uPDF
as in Tab.~\protect\ref{phi}. The lower plots always show the 
ratio
$R=\frac{theory-data}{data}$. }
\label{fig-phi-zeus}
\end{wrapfigure}

In Fig.~\ref{fig-phi-zeus} we show a comparison of the measurement of \cite{zeus-delta-phi}
with the prediction of CASCADE using the uPDF determined before. A reasonable
description of the measurement is achieved. Table~\ref{phi}
 shows the $\chi^{2}/ndf$ obtained for these data
and also to the azimuthal correlations from \cite{h1-delta-phi}.
\begin{wraptable}{l}{0.8\columnwidth}
\centerline{\begin{tabular}{|l|c|c|c|}
\hline
 & & \multicolumn{2}{c|}{$\chi^2/ndf$} \\\hline
Bg &$\mu$ [GeV]  & $\frac{d\sigma}{dQ^2 d\Delta \phi}$ (H1 prel)\protect\cite{h1-delta-phi}
 &$\frac{d\sigma}{d\Delta \phi}$ (dijets ZEUS)\protect\cite{zeus-delta-phi}  \\\hline
 0.025  &  1.5    &  163/29=5.6  & 332/19=17.5   \\ \hline
 0.25   &  1.5    &  128/29=4.4  &  234/19=12.3  \\ \hline
 0.025  &  0   &  200/29=6.9  &  417/19=22.0  \\ \hline
 0.25   &  0    & 237/29=8.2   & 338/19=17.8    \\ \hline
\end{tabular}}
\caption{Quality of the description of $\frac{d\sigma}{ d\Delta \phi}$ 
using $Bg=0.025$ and $Bg=0.25$ together with $\sigma=1.5$~GeV. }
\label{phi}
\end{wraptable}
It is interesting to observe, that $\frac{d\sigma}{ d\Delta \phi}$ gives also 
access to $Bg$, now with a 
preference to a much steeper initial gluon distribution. The 
measurement  prefers a distribution which 
decreases for very small transverse momenta $\kt$. However it should be noted, that 
the form of the intrinsic $\kt$ distribution is not constrained.
\section{Conclusion}
The shape of the starting gluon distribution in $x$ and $\kt$ has been investigated
with dijet events in DIS. Whereas the cross sections as a function of
$E_t$ prefer a soft gluon distribution ($Bg \sim 0.025$) and show little 
sensitivity to the intrinsic $\kt$ distribution, the cross sections as a function 
of $ \Delta \phi$ prefer a much steeper gluon ($Bg \sim 0.25$) and show a clear
preference to a intrinsic $\kt$ distribution which decreases for small $\kt$.
The different $x$-slope of the initial gluon distribution, as already observed in 
fits to $F_2$ and $F_2^c$, is also observed in di-jet cross section measurement.
Further investigations are obviously needed. 

\section*{Acknowledgments}
Many thanks go to the
the organizers of this very interesting workshop.

\raggedright
\begin{footnotesize}

\end{footnotesize}


\end{document}